\documentclass[twocolumn,dvipsnames,onecolappendix]{aastex631}
\usepackage{amsmath,amsfonts,amssymb,graphicx,multirow}
\usepackage{mathrsfs}
\usepackage[]{hyperref}
\hypersetup{colorlinks=true}

\usepackage{booktabs}
\usepackage{amsmath}

\begin{document}

\received{Feb. 18, 2026}
\revised{Mar. 5, 2026}
\accepted{Mar. 10, 2026}

\title{A positive period derivative in the quasi-periodic eruptions of ZTF19acnskyy}

\author[0000-0002-0568-6000]{Joheen Chakraborty}
\thanks{joheen@mit.edu}
\author[0000-0003-3182-5569]{Saul A. Rappaport}
\affiliation{Department of Physics \& Kavli Institute for Astrophysics and Space Research, Massachusetts Institute of Technology, Cambridge, MA 02139, USA}

\author[0000-0003-4054-7978]{Riccardo Arcodia}
\affiliation{Department of Physics \& Kavli Institute for Astrophysics and Space Research, Massachusetts Institute of Technology, Cambridge, MA 02139, USA}
\affiliation{Black Hole Initiative at Harvard University, 20 Garden Street, Cambridge, MA 02138, USA}

\author[0000-0002-8304-1988]{Itai Linial}\thanks{NASA Einstein Fellow}
\affiliation{Department of Physics \& Columbia Astrophysics Laboratory, Columbia University, New York, NY 10027, USA}
\affiliation{Center for Cosmology and Particle Physics, Physics Department, New York University, New York, NY 10003, USA}

\author[0000-0003-0707-4531]{Giovanni Miniutti}
\affiliation{Centro de Astrobiolog\'ia (CAB), CSIC-INTA, Camino Bajo del Castillo s/n, 28692 Villanueva de la Ca\~nada, Madrid, Spain}

\author[0000-0002-7226-836X]{Kevin B. Burdge}
\affiliation{Department of Physics \& Kavli Institute for Astrophysics and Space Research, Massachusetts Institute of Technology, Cambridge, MA 02139, USA}

\author[0000-0003-1965-3346]{Jorge Cuadra}
\affiliation{Departamento de Ciencias, Facultad de Artes Liberales,Universidad Adolfo Ibáñez, Av.\ Padre Hurtado 750, Viña del Mar, Chile}
\affiliation{Millennium Nucleus on Transversal Research and Technology to Explore Supermassive Black Holes (TITANS), Gran Breta\~na 1111, Playa Ancha, Valpara\'iso, Chile}

\author[0000-0002-1329-658X]{Margherita Giustini}
\affiliation{Centro de Astrobiolog\'ia (CAB), CSIC-INTA, Camino Bajo del Castillo s/n, 28692 Villanueva de la Ca\~nada, Madrid, Spain}

\author[0000-0002-8606-6961]{Lorena Hernández-García}
\affiliation{Instituto de Estudios Astrof\'isicos, Facultad de Ingenier\'ia y Ciencias, Universidad Diego Portales, Av. Ej\'ercito Libertador 441, Santiago, Chile}
\affiliation{Centro Interdisciplinario de Data Science, Facultad de Ingenier\'ia y Ciencias, Universidad Diego Portales, Av. Ej\'ercito Libertador 441, Santiago, Chile}

\author[0000-0003-0172-0854]{Erin Kara}
\affiliation{Department of Physics \& Kavli Institute for Astrophysics and Space Research, Massachusetts Institute of Technology, Cambridge, MA 02139, USA}

\author[0000-0001-5231-2645]{Claudio Ricci}
\affiliation{Department of Astronomy, University of Geneva, ch. d’Ecogia 16, 1290, Versoix, Switzerland}
\affiliation{Instituto de Estudios Astrof\'isicos, Facultad de Ingenier\'ia y Ciencias, Universidad Diego Portales, Av. Ej\'ercito Libertador 441, Santiago, Chile}

\author[0000-0003-0820-4692]{Paula Sánchez-Sáez}
\affiliation{European Southern Observatory, Karl-Schwarzschild-Strasse 2, 85748 Garching bei München, Germany}

\author[0000-0003-3024-7218]{Philippe Yao}
\affiliation{Department of Astrophysical Sciences, Princeton University, Princeton, NJ 08544, USA}

\begin{abstract}
We report the first direct measurement of the period derivative in a quasi-periodic eruption (QPE), finding a smoothly \textit{increasing} period with $\dot{P}\approx (1.7\pm 0.02)\times10^{-2}$ d d$^{-1}$ in the source ZTF19acnskyy/``Ansky''. Most models for QPEs invoke repeated interactions of a stellar-mass orbiting companion around the supermassive black hole (SMBH) in an extreme mass-ratio inspiral (EMRI). In these scenarios, a positive $\dot{P}$ is surprising, but not impossible to produce. We explore several possible explanations for the observed $\dot{P}$, including stable mass-transfer driven by impulsive mass loss events in an EMRI, velocity kicks at pericenter due to tidal interactions with the SMBH, apparent period changes due either to general relativistic precession effects in an EMRI or light travel-time delays in a hierarchical SMBH binary, and mass-transfer variations in a thermal/viscous disk instability model. We find that none of the considered models provides a complete explanation for the data, motivating further work on physical explanations for positive period derivatives in QPEs.
\end{abstract}

\keywords{Supermassive black holes (1663); X-ray astronomy (1810); Transient sources (1851)}

\section{Introduction} \label{sec:intro}
Quasi-periodic eruptions (QPEs) are recurring soft X-ray transients from the supermassive black holes (SMBHs) in some nearby galaxy nuclei \citep{Miniutti19,Giustini20,Arcodia21,Arcodia24a,Arcodia25,Chakraborty21,Chakraborty25a,Quintin23,Nicholl24,Hernandez25a,Baldini26}. They show peak luminosities ranging from $L_{\rm peak}\sim 10^{42-44}$ erg s$^{-1}$, recurrence times of $T_{\rm rec}\sim 2.5-300\;$hr, blackbody-like spectra with temperatures of $kT\sim 50-250\,$ eV, SMBH masses of $\sim 10^{5-7.5}M_\odot$, and host galaxy redshifts up to $z\lesssim 0.1$. In some sources, QPEs have emerged months to years after Tidal Disruption Events (TDEs) selected by X-ray \citep{Miniutti23a,Chakraborty21,Baldini26} and optical surveys \citep{Quintin23,Nicholl24,Chakraborty25a}, providing an interesting theoretical puzzle as well as a promising avenue for future discoveries. 

The leading model class thus far explains the recurring X-ray bursts via repeated interactions with a gravitationally captured stellar-mass object around the SMBH in an extreme mass-ratio inspiral (EMRI; e.g.~\citealt{King20, Xian21,Sukova21,Krolik22,Linial23a,Lu23,Franchini23,Linial23b,Tagawa23,Yao25a})---though we will soon see the more appropriate term in our case may be extreme mass-ratio \textit{outspiral}. In this picture, the bursts may arise from collisions of the orbiting companion with the accretion disk surrounding the SMBH (perhaps initially formed by the precursor TDE), or by repeated mass-loss of the companion near pericenter. EMRI models have provided an attractive framework to interpret the zeroth-order observational properties of QPEs, such as their amplitudes, quasi-periodicity, rates, and spectra. Still, there remain multiple open questions related to e.g.~the precise source of short- and long-term timing variations observed in some QPEs \citep{Miniutti25,Arcodia26} and the burst energetics across the population \citep{Mummery25,Linial25}. Alternative models of accretion disk instabilities driven by e.g.~nodal precession \citep{Raj21,Middleton25} or variations in magnetic viscosity \citep{Sniegowska20,Sniegowska23,Pan22,Pan23,Pan25,Kaur23} are thus worthwhile avenues for ongoing investigation. It is entirely possible that the growing family of QPEs may arise from a combination of different physical mechanisms.

The most extreme known source of QPEs is the nuclear transient ZTF19acnskyy/``Ansky'' \citep{Sanchez24}, which hosts long, luminous eruptions of integrated energy budget $L\Delta t\gtrsim 100\times$ larger and recurrence time $\gtrsim 10\times$ longer than typical QPE sources \citep{Hernandez25a}. The eruptions also uniquely show broad, time-evolving absorption lines potentially indicating they are powered by relativistic outflows \citep{Chakraborty25b}. The nature of the optical/UV transient preceding the X-ray QPEs in Ansky is uncertain, with possible interpretations including an AGN turn-on around a $\sim 10^6M_\odot$ SMBH or a featureless TDE \citep{Sanchez24,Zhu25}. Ansky is the only QPE currently known to show a UV counterpart to the X-ray bursts \citep{Guo26}, possibly as a result of the long timescales and large energy budget associated with the flares \citep{Vurm24,Chakraborty25b}. Furthermore, \cite{Hernandez25b} recently reported a \textit{doubled} recurrence time and burst duration between 2024-2025, followed by a steady increase of $\sim 0.1$ d per flare. Ansky's extraordinary characteristics provide a stress-test for the very limits of any physical model of QPEs.

In this paper, we present data from our continued X-ray monitoring campaign throughout 2025-2026. In total, we observed 23 bursts---of which 19 were consecutive, the most in any source thus far---which enable a unique timing analysis sensitive to secular period evolution. We outline the observations and data reduction procedures in Section~\ref{sec:methods} and present the key results of our timing analysis in Section~\ref{sec:results}. We consider several models for the increasing period, and discuss their implications for the long-term fate of the QPEs in Ansky, in Section~\ref{sec:discussion}. We make concluding remarks in Section~\ref{sec:conclusion}.

\section{Observations and methods} \label{sec:methods}
\begin{figure*}
    \centering
    \includegraphics[width=\linewidth]{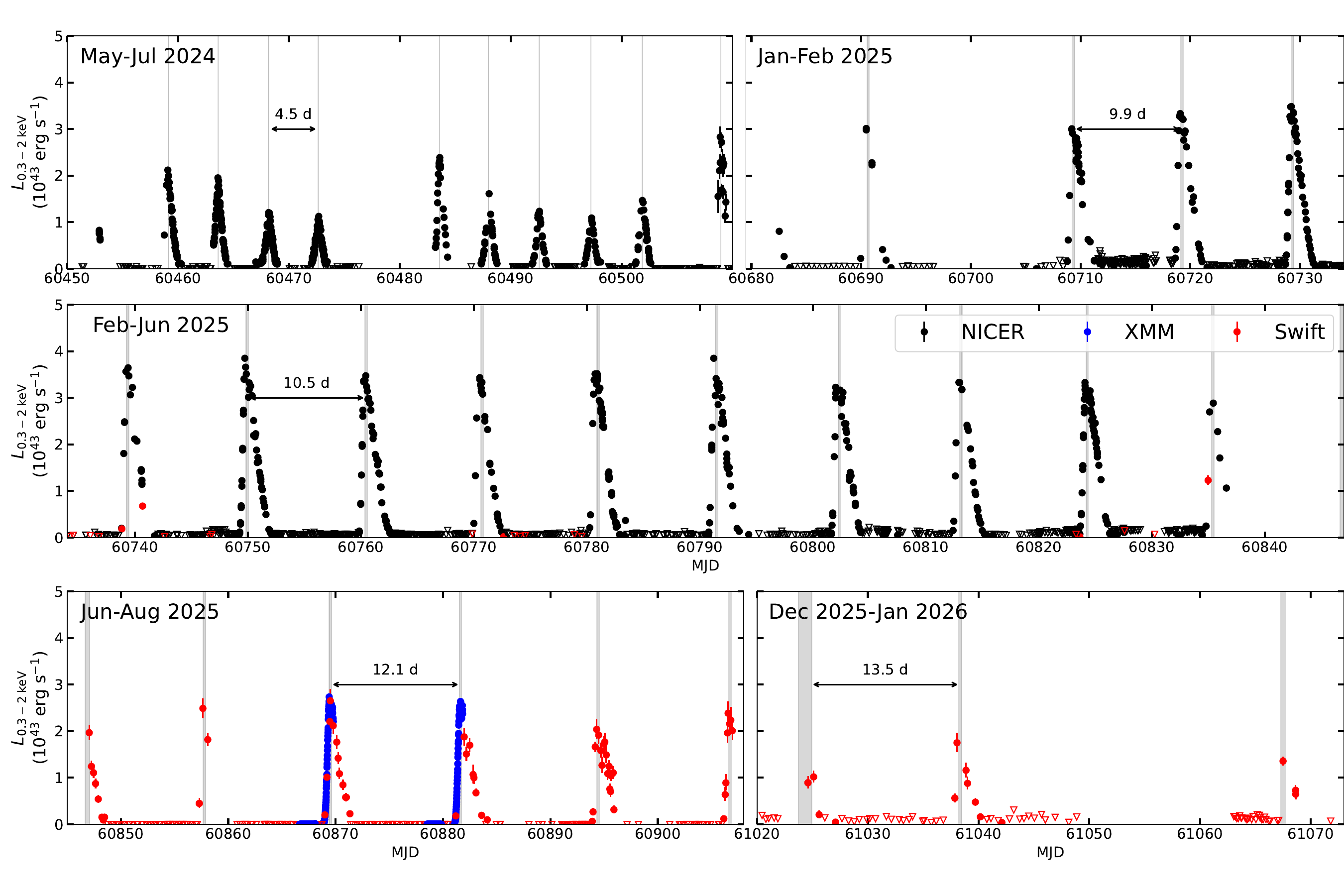}
    \caption{The soft X-ray light curve of Ansky from 2024-2026. Data were taken with the \textit{NICER} (black), \textit{Swift} (red), and \textit{XMM-Newton} (blue) observatories. Gray shaded regions correspond to the $\pm1\sigma$ contours of $t_{\rm peak}$ measurements from our timing model (Table~\ref{tab:peak_times}).}
    \label{fig:lc}
\end{figure*}

\begin{table}
\centering
\caption{Observed peak times and luminosities throughout 2025-2026. Measurement errors are added in quadrature with a systematic model timing uncertainty of $0.1$ d.}
\label{tab:peak_times}
\begin{tabular}{ccccc}
\toprule
Epoch & MJD & $\sigma_-$ & $\sigma_+$ & $L_{\rm peak}$  \\
& & (days) & (days) & ($10^{43}$ erg s$^{-1}$) \\
\midrule
-2 & 60690.63 & 0.10 & 0.10 & $4.2^{+0.5}_{-0.5}$ \\
\specialrule{0.1pt}{0.3em}{0.3em}
0 & 60709.36 & 0.10 & 0.10 & $3.6^{+0.4}_{-0.4}$ \\
1 & 60719.23 & 0.10 & 0.10 & $4.3^{+0.5}_{-0.5}$ \\
2 & 60729.32 & 0.10 & 0.10 & $4.2^{+0.4}_{-0.4}$ \\
3 & 60739.35 & 0.10 & 0.10 & $5.7^{+0.6}_{-0.6}$ \\
4 & 60749.94 & 0.10 & 0.10 & $6.0^{+0.6}_{-0.6}$ \\
5 & 60760.46 & 0.10 & 0.10 & $5.3^{+0.5}_{-0.5}$ \\
6 & 60770.71 & 0.10 & 0.10 & $4.6^{+0.5}_{-0.5}$ \\
7 & 60781.00 & 0.10 & 0.10 & $4.6^{+0.5}_{-0.5}$ \\
8 & 60791.46 & 0.10 & 0.10 & $4.6^{+0.5}_{-0.5}$ \\
9 & 60802.32 & 0.10 & 0.10 & $4.2^{+0.4}_{-0.4}$ \\
10 & 60813.09 & 0.10 & 0.10 & $4.8^{+0.5}_{-0.5}$ \\
11 & 60824.27 & 0.10 & 0.10 & $4.5^{+0.5}_{-0.5}$ \\
12 & 60835.38 & 0.11 & 0.11 & $3.2^{+0.4}_{-0.4}$ \\
13 & 60846.83 & 0.17 & 0.24 & $3.8^{+1.6}_{-1.6}$ \\
14 & 60857.78 & 0.14 & 0.12 & $3.1^{+1.1}_{-0.7}$ \\
15 & 60869.51 & 0.10 & 0.10 & $2.8^{+0.3}_{-0.3}$ \\
16 & 60881.64 & 0.10 & 0.10 & $2.6^{+0.3}_{-0.3}$ \\
17 & 60894.43 & 0.11 & 0.11 & $2.3^{+0.3}_{-0.3}$ \\
18 & 60906.71 & 0.12 & 0.13 & $2.5^{+0.3}_{-0.4}$ \\
\specialrule{0.1pt}{0.3em}{0.3em}
27 & 61024.83 & 1.08 & 0.10 & $2.2^{+1.3}_{-0.7}$ \\
28 & 61038.33 & 0.14 & 0.13 & $2.1^{+0.5}_{-0.5}$ \\
\specialrule{0.1pt}{0.3em}{0.3em}
30 & 61067.48 & 0.22 & 0.19 & $2.0^{+1.1}_{-0.8}$ \\
\bottomrule
\end{tabular}
\end{table}

The \textit{NICER} X-ray Timing Instrument \citep{Gendreau16} aboard the International Space Station subsequently observed Ansky for a total of 282.5~ks across 263 observations from Jan 05-Jun 16, 2025. We followed the time-resolved spectroscopy approach for light curve estimation outlined in Section 2.1 of \cite{Chakraborty24}. Spectral fitting and background estimation were performed with the \texttt{SCORPEON} model over a broadband energy range (0.25--10 keV) for data taken in orbit night, and a restricted range (0.38--10 keV) during orbit day. We grouped our spectra with the optimal binning scheme of \cite{Kaastra16} (\texttt{grouptype=optmin} with \texttt{groupscale=10} in the \texttt{ftgrouppha} command) and performed all spectral fitting with the Cash statistic \citep{Cash1979}.

\textit{Swift} X-ray telescope (XRT) data were obtained from the online interface (\href{https://www.swift.ac.uk/LSXPS}{https://www.swift.ac.uk/LSXPS}) to the Living \textit{Swift} XRT Point Source (LSXPS) Catalogue \citep{Evans23a}. LSXPS is an automatically updated repository of all \textit{Swift} XRT observations of $>100$ second duration in Photon Counting (PC) mode. LSXPS reports X-ray fluxes in counts per second (cps); we converted to cgs units using a conversion factor of $1\;\rm cps=2.2\times 10^{-11}$ erg cm$^{-2}$ s$^{-1}$, which was computed for a 100 eV blackbody spectrum and Galactic neutral absorption with $N_{\rm H}=2.6\times 10^{20}$ cm$^{-2}$.

\textit{XMM-Newton} data were obtained through GO 096454 in AO 24 (PI: Chakraborty). The observations were taken on July 10/12/22/24, 2025 (OBSIDs 0964540101-0964540401), with the first and third observing quiescence and the second and fourth observing the rise-to-peak of two consecutive eruptions. The \textit{XMM} data provide a deep view into the short-timescale X-ray spectral-timing variability, and will be the subject of forthcoming work. The data were reduced using \textit{XMM} SASv21.0.0 and HEASoft v6.33. Source products were extracted from a circular region of 33$''$ radius, while the background was extracted from source-free circular region falling on the same detector with a 60$''$ radius. We retained events with PATTERN$\leq$4 (single and double events only) and discarded time intervals with a 10--12 keV count rate $\geq 1$ counts s$^{-1}$. Light curves were extracted with \texttt{evselect}, corrected for detector efficiency, vignetting, PSF, and bad pixels using \texttt{epiclccorr}, and binned to 1~ks in the 0.3-2 keV band.

\begin{figure}
    \centering
    \includegraphics[width=\linewidth]{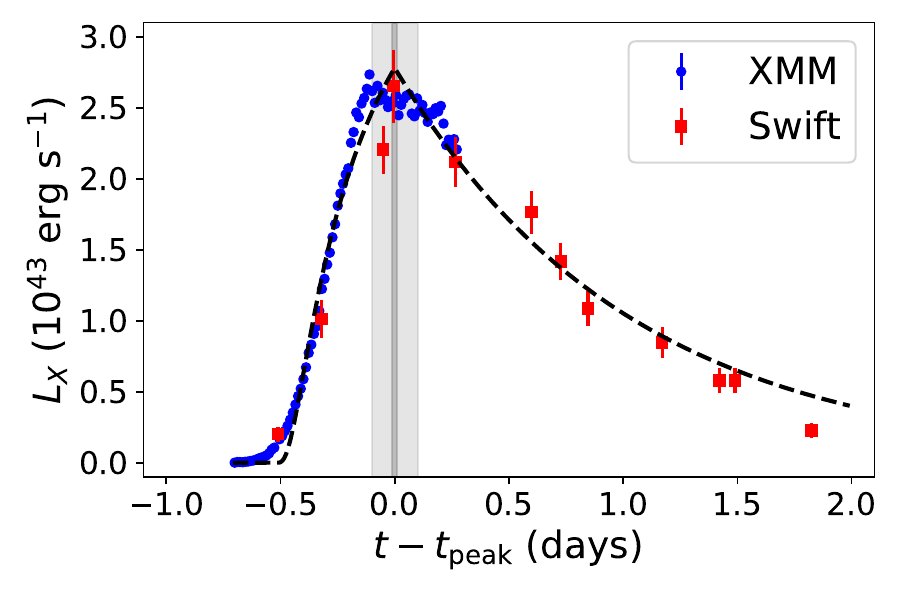}
    \caption{Example burst profile from an \textit{XMM}-observed burst (Epoch 15, $t_{\rm peak}=60869.51$). The dark gray band shows the model $t_{\rm peak}$ error estimated via MCMC. The light gray band shows our additional 0.1 day systematic uncertainty, which accounts for the scatter in \textit{XMM} and \textit{Swift} data near the peak.}
    \label{fig:burst_profile}
\end{figure}

\section{Results} \label{sec:results}

In Fig.~\ref{fig:lc} we show the full light curve of the QPEs in Ansky. To estimate the peak timings, we fit each eruption with a four-parameter exponential rise-and-decay model introduced for QPEs in \citealt{Arcodia22}:
\begin{equation*}
    L(t) = \begin{cases}
        L_{\rm peak}\lambda e^{\tau_1/(t_{\rm peak}-t_{\rm as}-t)} & \mathrm{if\;}t\leq t_{\rm peak}\\
        L_{\rm peak}e^{-(t-t_{\rm peak})/\tau_2} & \mathrm{if\;}t>t_{\rm peak}
    \end{cases}
\end{equation*}
where $L_{\rm peak}$ is the flare amplitude, $t_{\rm peak}$ is the peak timing, and $\tau_1$/$\tau_2$ are the rise/decay $e$-folding times. $\lambda\equiv \exp(\sqrt{\tau_1/\tau_2})$ and $t_{\rm as}\equiv\sqrt{\tau_1\tau_2}$ are derived parameters setting the normalization and asymptote time; for more discussion, see \cite{Arcodia22} or \cite{Chakraborty24}. Uncertainties on $t_{\rm peak}$ were initially estimated via MCMC, resulting in characteristic errors of $\sim 0.01$ d. However, in several cases we found the best-fit model was not a perfect description of the data, and that the model $t_{\rm peak}$ disagreed with the measured peak flux by up to $\lesssim 0.1$ days (see Fig.~\ref{fig:burst_profile} for an illustrative example of the burst with highest-cadence coverage); as a result, we added in quadrature a 0.1 d uncertainty to all MCMC error estimates to capture the intrinsic model inaccuracy in describing the peak timings. We verified that the choice of $\sigma_{\rm sys}\in\{0.05,0.1,0.15,0.2\}$d does not qualitatively alter our results. The model peak times and luminosities are reported in Table~\ref{tab:peak_times}.

The bursts reach $L_{\rm peak} \approx (2-6)\times 10^{43}$ erg s$^{-1}$, corresponding to time-integrated energy outputs $L\Delta t\approx (1-5)\times10^{48}$ erg. Assuming radiative efficiencies in the range $\eta\sim 0.01-1$, we find characteristic mass budgets of $\Delta M_*\sim L\Delta t/(\eta c^2) \sim 10^{-6}-10^{-4}M_\odot$ for accretion-powered emission, or $\Delta M_*\sim 10^{-3}-10^{-1}M_\odot$ for emission powered by shocks at velocities comparable to Keplerian for a circular orbit ($v_{\rm K}\sim0.03c$). Given that $\mathcal{O}(100)$ bursts are expected to have occurred thus far---including those observed in 2024 and 2026, and interpolating between observational gaps---we may infer a total mass budget on the order of $10^{-4}-10^{-2}M_\odot$ for accretion-powered QPEs, or $0.1-10M_\odot$ for collisional shock-powered QPEs. The former model thus remains consistent with being powered by a stellar-mass object, while the latter may strain the total mass budget available within the stellar EMRI model (though see~\citealt{Linial25}).

\begin{figure*}
    \centering
    \includegraphics[width=\linewidth]{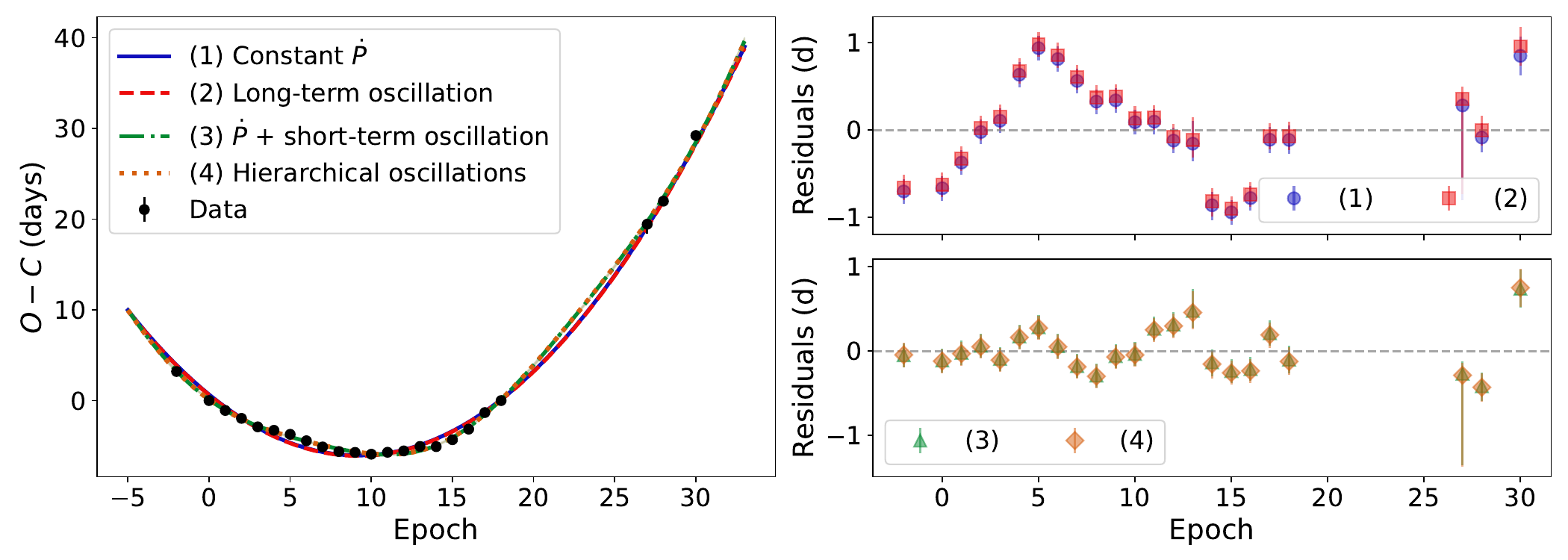}
        \caption{\textbf{Left:} $O-C$ diagram of the QPEs overplotted with models 1-4. The data are consistent with $P_0\approx 9.5$ d and $\dot{P}\approx 1.7\times10^{-2}$ d d$^{-1}$ at $T_0\approx 60710.1$. In principle, it is possible that the apparent period increase is due to a long-term period oscillation observed locally (models 2 and 4); the $>1$ yr data baseline over which $\dot{P}$ is constant constrains any such oscillation to have a period $\gtrsim 11$ yr and amplitude $\gtrsim 1000$ d. \textbf{Right:} data$-$model residuals. Models 1-2 show structured residuals, which models 3-4 interpret as a sine with period $\sim 155$ d and amplitude $\sim 0.8$ d (though see Appendix~\ref{app:oc_noise}).}
    \label{fig:oc}
\end{figure*}

In Fig.~\ref{fig:oc} we show an $O-C$ (observed minus calculated) diagram of the 2025 QPEs, where $O$ are the observed peak arrival times (Table~\ref{tab:peak_times}) and $C$ are the calculated timings assuming a fixed period, i.e. $C=T_0+nP_0$ at $n$ epochs after a reference timing $T_0$ for a trial period $P_0$. The dominant feature visible in the $O-C$ diagram is a concave-up trend over the $\sim 30$ observed epochs. We consider four possible models to describe the timings:
\begin{align}
T_{\rm M,1} &= T_0 + \Delta P_0n + \frac{1}{2}P_0\dot{P}n^2 + \cdots \\
T_{\rm M,2} &= A_l\sin\Big(\frac{2\pi P_0}{P_l}n+\phi_l\Big) + T_0 + \Delta P_0n \\
T_{\rm M,3} &= A_s\sin\Big(\frac{2\pi P_0}{P_s}n+\phi_s\Big) + T_0 + \Delta P_0n + \frac{1}{2}P_0\dot{P}n^2 + \cdots \\
T_{\rm M,4} &= A_l\sin\Big(\frac{2\pi P_0}{P_l}n+\phi_l\Big) \nonumber \\
&\phantom{=}+ A_s\sin\Big(\frac{2\pi P_0}{P_s}n+\phi_s\Big) + T_0 + \Delta P_0n
\end{align} \\
where $T_{\mathrm{M},N}$ is the model timing of the $n$th epoch of model $N$; $\dot{P}$ is the period time-derivative; $\Delta P_0$ is a correction to the trial period $P_0$; and $A_{l/s}$, $P_{l/s}$, and $\phi_{l/s}$ are the amplitude, period, and phase of the long/short-period sinusoidal components in models 2-4. All four models and their residuals are plotted in Fig.~\ref{fig:oc}, with best-fitting parameters given in Table~\ref{tab:oc_params}.

\begin{figure}
    \includegraphics[width=\linewidth]{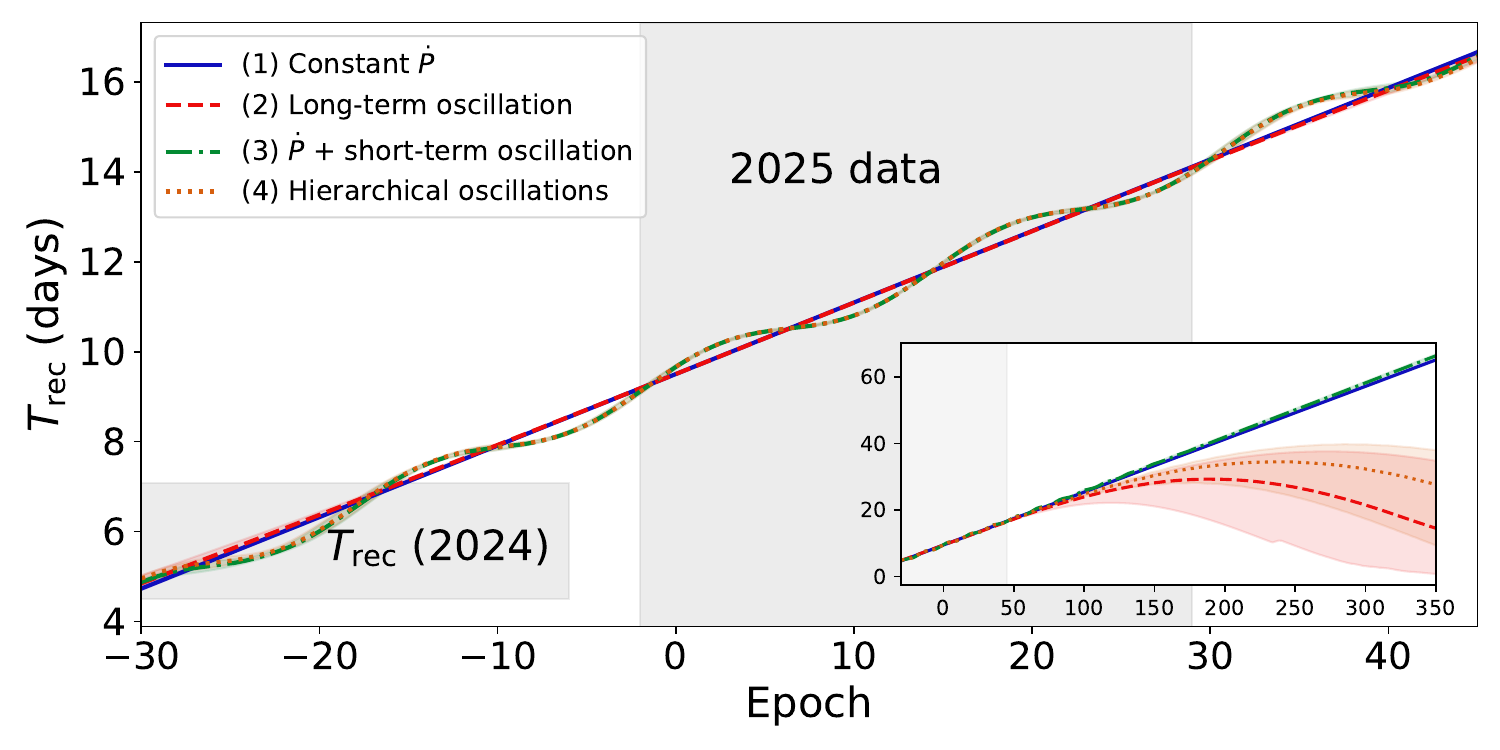}
    \centering
        \caption{We plot the period evolution for the four $O-C$ models. Shaded regions denote the 2025 data (Epochs -2-29) and the range of periods observed in 2024. The inset panel extrapolates over $\sim$few hundred epochs, showing the constant $\dot{P}$ and long-term oscillation models begin to diverge after $\gtrsim100$ epochs.}
    \label{fig:period_extrapolation}
\end{figure}

Models 1 and 3 involve a series expansion of the QPE period about $P_0$, carried out to quadratic order (a constant $\dot{P}$). Models 2 and 4 consist only of sinusoidal modulation(s) of the burst arrival times, as is seen increasingly commonly in QPE timing analyses \citep{Chakraborty24,Miniutti25,Arcodia26}. We note that for $n\ll P_l/P_0$, a Taylor expansion of the long-period sinusoidal component in models 2 and 4 yields a quadratic term $\propto n^2$, making models 1 and 2 (and 3 and 4) mathematically identical at quadratic order; the models become clearly distinguishable only after $\gtrsim 100$ further epochs (Fig.~\ref{fig:period_extrapolation}).

While the absolute values of $\chi^2$ are meaningless given our ad hoc choice of a $0.1$ d systematic error, the difference in reduced chi-squared can be used for qualitative model comparison. Models 1/2/3/4 have $\chi^2/\mathrm{dof.} = 14.9/15.9/2.9/3.1$ respectively. Models 3 and 4 are thus favored over 1 and 2, while model 3---a period derivative plus a short-term sinusoidal modulation---achieves the best fit statistic.

\begin{table*}
\centering
\caption{Best-fit parameters for the four $O-C$ timing models. Dashes indicate parameters not present in a given model.}
\label{tab:oc_params}
\begin{tabular}{lcccc}
\toprule
Parameter & Model 1 & Model 2 & Model 3 & Model 4 \\
\midrule
$T_0$ (MJD) & $60710.02 \pm 0.06$ & --- & $60710.09^{+0.09}_{-0.10}$ & --- \\
$P_0$ (d) & $9.50 \pm 0.01$ & --- & $9.48^{+0.02}_{-0.02}$ & --- \\
$\dot{P}$ (d d$^{-1}$) & $1.67^{+0.01}_{-0.01} \times 10^{-2}$ & --- & $1.72^{+0.01}_{-0.02} \times 10^{-2}$ & --- \\
\specialrule{0.1pt}{0.3em}{0.3em}
$A_l$ (d) & --- & $1872^{+2309}_{-1077}$ & --- & $3068^{+1498}_{-1205}$ \\
$P_l/P_0$ (cyc) & --- & $681^{+338}_{-238}$ & --- & $863^{+190}_{-190}$ \\
$\phi_l$ (rad) & --- & $4.62^{+0.03}_{-0.04}$ & --- & $4.65^{+0.01}_{-0.02}$ \\
\specialrule{0.1pt}{0.3em}{0.3em}
$A_s$ (d) & --- & --- & $0.79 \pm 0.05$ & $0.79 \pm 0.05$ \\
$P_s/P_0$ (cyc) & --- & --- & $16.2^{+0.5}_{-0.5}$ & $16.3^{+0.6}_{-0.6}$ \\
$\phi_s$ (rad) & --- & --- & $-0.91^{+0.15}_{-0.13}$ & $-0.91 \pm 0.15$ \\
\specialrule{0.1pt}{0.3em}{0.3em}
$\chi^2$/dof & 297.7/20 & 302.1/19 & 49.3/17 & 49.6/16 \\
$\chi^2_\nu$ & 14.9 & 15.9 & 2.9 & 3.1 \\
\bottomrule
\end{tabular}
\end{table*}

\textbf{Period derivative:} models 1 and 3 interpret the concave-up trend as a secular period increase expanded to quadratic order. The best-fitting model 1 corresponds to $P_0=9.50\pm0.01$ d and $\dot{P}= (1.67\pm 0.01)\times10^{-2}$ d d$^{-1}$ at $T_0=60710.02\pm0.06$ MJD, whereas in model 3 it is $P_0=9.48\pm0.02$ d and $\dot{P}=(1.72\pm 0.02)\times10^{-2}$ d d$^{-1}$ at $T_0=60710.10\pm0.09$ MJD.

\textbf{Long-period oscillation:} models 2 and 4 instead interpret the period increase as part of a long-term oscillation, and thus that the period will eventually decrease. This sinusoidal component in model 2 corresponds to an amplitude of $A_l=1872^{+2309}_{-1077}$ d with a period of $P_l/P_0=681^{+338}_{-238}$ (or $P_l\approx 17.7^{+8.8}_{-6.2}$ yr). In model 4, the amplitude is $A_l=3068^{+1498}_{-1205}$ d with a period of $P_l/P_0=863\pm 190$ (or $P_l\approx 22.4\pm4.9$ yr).

\textbf{Short-period oscillation:} motivated by the apparent sine-like residuals in models 1-2, models 3 and 4 contain an additional short-term modulation. In model 3 it has an amplitude of $A_s=0.79\pm 0.05$ d and $P_s/P_0=16.2\pm0.5$ (or $P_s\approx 154$ d). In model 4 the amplitude is $A_s=0.79\pm 0.05$ d, with period $P_s/P_0=16.3\pm 0.6$ (or $P_s\approx 155$ d).

In summary, the data are consistent with a constant period derivative with $P_0\approx 9.5\pm0.02$ d and $\dot{P}\approx(1.7\pm0.02)\times10^{-2}$ d d$^{-1}$ at $T_0\approx 60710.1$. This can instead be interpreted as a long-term period oscillation, in which case the period is $\approx 11-27$ yr and the amplitude is $\approx 1000-4500$ d. We also fit an additional shorter-period oscillation with $A_s=0.79\pm0.05$ d and period $\approx 155$ d, though we caution that the 19 consecutive epochs span only $\sim$1.2 cycles, and that the reality of this secondary oscillation depends on the physical interpretation of the QPE-generating process (Appendix~\ref{app:oc_noise}).

We also note that the 2024 data showed recurrence times between $\sim 4.5-7$ days (Fig. 1; \citealt{Hernandez25a}), which is considerably shorter than the $P_0$ inferred from the 2025 data. However, given the long observation gap between 2024 and 2025, it is impossible to unambiguously assign epoch numbers to the 2024 data, so we cannot include them in the $O-C$ analysis. Still, extrapolating the period backwards finds recurrence times plausibly consistent with the 2024 observations at epochs $\lesssim-20$ (Fig.~\ref{fig:period_extrapolation}), providing a convenient explanation for those data. The inset panel of Fig.~\ref{fig:period_extrapolation} extrapolates the period forwards, finding that the constant $\dot{P}$ and long-term oscillation models begin to diverge at epochs $\gtrsim 100$. 

\section{Discussion} \label{sec:discussion}

Given the formal degeneracy between a constant $\dot{P}\approx1.7\times10^{-2}$ d d$^{-1}$ and the leading-order Taylor expansion of a long-period oscillation over $11-27$ yr with the current $\sim 30$-epoch baseline, as well as the variety of physical processes which may be relevant to period evolution in repeating nuclear transients \citep{Linial24a}, we consider several models for the observed period change. In Sections~\ref{subsec:stable_mdot}-\ref{subsec:pericenter_kicks} we assume QPEs are powered by repeatedly stripping mass from a bound star near pericenter (akin to a repeating TDE, though not necessarily accretion-powered), which results in a steady increase of the orbital period as in models 1 \& 3. In Sections~\ref{subsec:gr_precession}-\ref{subsec:third_body} we assume there is no secular change to the orbital period, but that the apparent $\dot{P}$ is the result of geometric light travel-time delays induced by general relativistic precession or the effect of a third body, à la models 2 \& 4. Section~\ref{subsec:disk_instability} discusses the ability of disk instability models to describe the data. 

\subsection{Orbital evolution due to mass-transfer} \label{subsec:stable_mdot}

Mass transfer from a lower-mass donor to a larger-mass companion can counteract angular momentum-loss mechanisms such as gravitational-wave emission or tidal dissipation and cause binary orbits to expand, an effect which has even been known theoretically for decades (see e.g.~\citealt{Eggleton}) and even measured directly in a few ultracompact binary systems via eclipse timing \citep{deMiguel18,Chakraborty24b}. This result has been extended to eccentric mass-transfering stellar binaries by deriving equations for the time evolution of the osculating orbital elements, e.g.~via the orbit-averaged Lagrange planetary equations under a perturbing force due to mass-transfer \citep{Hadjidemetriou63,Sepinsky09,Dosopoulou16a,Dosopoulou16b}. While the general case depends on the orbital angular momentum retention factor, size of the donor's Roche lobe, donor/accretor masses, and other uncertain system properties, most of these are linear-order in the mass ratio $q\equiv M_*/M_\bullet$, and can thus be dropped due to the extreme mass ratio in our system ($q\sim10^{-6}$). We follow Eqns.~43-44 of \cite{Dosopoulou16b}, which assume delta-function Roche lobe overflow at pericenter and non-conservative mass transfer ($\dot{M}_{\rm tot},\dot{J}_{\rm orb}\neq0$). We note that their formalism assumes mass is lost only through the inner Lagrange point L1, as is appropriate for stellar-mass binaries. In the extreme mass ratio limit ($q\sim10^{-6}$), L2 mass loss becomes nontrivial (\citealt{Linial17}; Section~\ref{subsec:pericenter_kicks}), which will alter the angular momentum transfer and thus the magnitude of $\Delta P_{\rm orb}/P_{\rm orb}$ by an uncertain degree. Still, as an order-of-magnitude estimate, we can take the $q\ll1$ limit to find the fractional change in orbital period ($P_{\rm orb}$):
\begin{equation}
    \frac{\Delta P_{\rm orb}}{P_{\rm orb}} \approx 3\zeta \bigg(\frac{\Delta M_*}{M_*}\bigg)\sqrt{1-e^2}
    \label{eq:delta_P}
\end{equation}
where $e$ is the orbital eccentricity; $\Delta M_*$ is the fraction of the stellar mass loss per passage; and $\zeta$ is the fraction of stripped mass captured by the primary, defined such that $\Delta M_\bullet=-\zeta\Delta M_*$, as opposed to being lost to infinity and carrying orbital angular momentum away with it (e.g.~as in \citealt{Linial17}). We note that Eq.~\ref{eq:delta_P} is reminiscent of the Newtonian result of \cite{Hughes19} obtained by enforcing adiabatic invariance of the orbital actions assuming conservative mass-transfer. Though the methods differ, they found $\Delta P_{\rm orb}/P_{\rm orb} = 3(\Delta M_*/M_*)$; our result differs by the order-unity factor $\sqrt{1-e^2}$ and the mass-retention term $\zeta$.

\begin{figure}
    \centering
    \includegraphics[width=\linewidth]{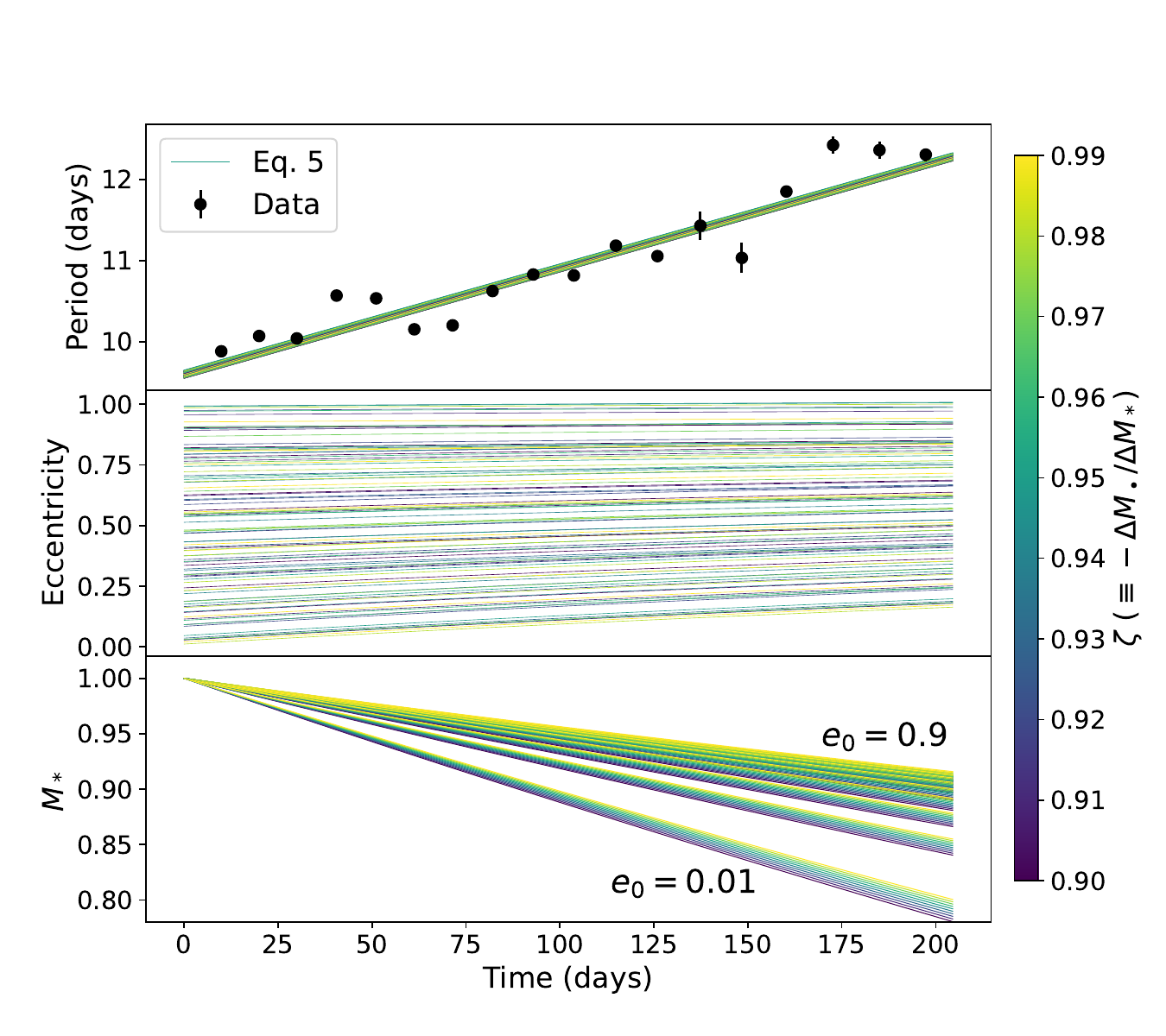}
    \caption{Evolution of $P_{\rm orb}$, $e$, and $M_*$ for various choices of ($e_0$, $\zeta$), assuming impulsive mass-loss occurs at pericenter, from Eqns.~\ref{eq:delta_P}-\ref{eq:delta_e}. Each trajectory on the top/middle panels has a random offset (up to 0.1/0.05 respectively) for visual clarity.}
    \label{fig:evol}
\end{figure}

Similarly, the change in orbital eccentricity is:
\begin{equation}
    \Delta e \approx 2\zeta\bigg(\frac{\Delta M_*}{M_*}\bigg)(1-e)\sqrt{1-e^2}
    \label{eq:delta_e}
\end{equation}
implying the eccentricity is excited to larger values as the semimajor axis increases. In Fig.~\ref{fig:evol} we plot the evolution of $P_{\rm orb}$, $M_*$, and $e$ according to Eqns.~\ref{eq:delta_P}-\ref{eq:delta_e} for various choices of ($e_0$, $\zeta$). The typical required fractional mass-loss per orbit is of order $\Delta M_*/M_*\approx (4-8)\times10^{-3}$, increasing for lower initial eccentricities (as evident from Eq.~\ref{eq:delta_P}); remarkably, this mass budget is comparable to values inferred from the energetics of the time-evolving outflow reported in Ansky \citep{Chakraborty25b}. The inferred fractional mass loss carries the dramatic implication that $\sim 10\%$ of the stellar mass must be stripped within just a year, placing a strict bound on the continued eruption lifetime and predicting dramatic luminosity evolution should be seen in continued monitoring.

The assumption of mass-loss from the orbiting star places some constraints on its nature. The Roche lobe at pericenter is approximately:
\begin{equation}
    R_L \approx 4.8 R_\odot\;\bigg(\frac{P_{\rm orb}}{11\;\rm d}\bigg)^{2/3}\bigg(\frac{1-e}{0.5}\bigg)\bigg(\frac{M_*}{M_\odot}\bigg)^{2/3}
    \label{eq:roche_lobe}
\end{equation}
meaning an orbiting companion with an outer radius of several $\times R_\odot$ and/or $e\gtrsim 0.9$ is required for Roche-lobe overflow. While assuming the main sequence mass-radius relation would imply this corresponds to a stellar mass of $\sim 7M_\odot$, recent work has found tidal heating in close proximity to the SMBH can lead to runaway radial expansion of the star \citep{Yao25b}, meaning even lower masses are plausible.

Dividing Eq.~\ref{eq:delta_P} by Eq.~\ref{eq:delta_e} and integrating implies that $r_p\equiv a(1-e)$ is a constant as $a$ and $e$ increase. This provides a convenient potential explanation for why the eruption luminosity and integrated energy per flare do not change dramatically throughout 2025 (Fig.~\ref{fig:lc}). One major uncertainty with this model is whether, after losing $\sim10\%$ of its mass within a year, the stellar structure would remain constant enough to power continued eruptions of comparable luminosity, even at a fixed pericenter distance. To this end, \cite{Bandopadhyay25} found that stars with masses $\gtrsim 1M_\odot$ become a factor of few \textit{denser} following the rapid stripping of 1-10\% their masses (see their Figs. 5, 6 \& 8), providing a circumstantial explanation for the lack of significant burst luminosity evolution (Fig.~\ref{fig:lc}).

Another uncertainty concerns the viscous timescale, which is approximately \citep{SS73}:
\begin{equation}
    t_{\rm visc} \sim 58\;\mathrm{days}\bigg(\frac{0.1}{\alpha}\bigg)\bigg(\frac{0.1}{H/R}\bigg)^2\bigg(\frac{M_\bullet}{10^6M_\odot}\bigg)\bigg(\frac{R}{100R_g}\bigg)^{3/2}
\end{equation}
In comparison, the 2025 eruptions have durations of just 3 days. Reconciling the small required radii $\lesssim 14 R_g$ (for $\alpha=0.1$ and $H/R=0.1$) with the orbital period and $M_\bullet$ would require an eccentricity of $e\approx 0.99$, precluding any stellar companion but a white dwarf (e.g. \citealt{King20}). Alternatively, a separate mechanism, such as collisional shocks of the stripped mass with the ambient disk, may be relevant for the system (e.g.~\citealt{Lu23}).

\subsection{Velocity kicks from tidal interaction with the SMBH} \label{subsec:pericenter_kicks}

The models of mass-transfer driven orbital evolution in Section~\ref{subsec:stable_mdot} are generally invoked for stellar-mass binaries, and do not consider the effects of the SMBH tidal field on the companion star. To this end, hydrodynamical simulations first carried out in \cite{Manukian13} found the surprising result that, in contradiction to the expectations from linear tidal theory, partially disrupted stars around SMBHs can receive kicks which eject them on progressively more \textit{unbound} orbits. This result was confirmed and extended by follow-up studies \citep{Gafton15,Cufari23,Chen24}, which generally attribute the positive-energy kicks to greater mass loss through L1 than L2. As mass flowing through L1 has a smaller orbital energy than the star, energy/momentum conservation can endow the stellar remnant with a positive-energy kick during its pericenter passage. This effect competes with deposition of orbital energy into tidal oscillatory modes of the star, which should decay the orbit; in general either may dominate, resulting in different signs of $\dot{P}$ in various regions of parameter space \citep{Ryu20,Chen24}. We also note that asymmetric mass loss is not definitively known to be the primary cause for this effect; see e.g.~\cite{Coughlin25} who propose an alternative mechanism related to the reformation of the stellar core after pericenter passage.

\begin{figure}
    \centering
    \includegraphics[width=\linewidth]{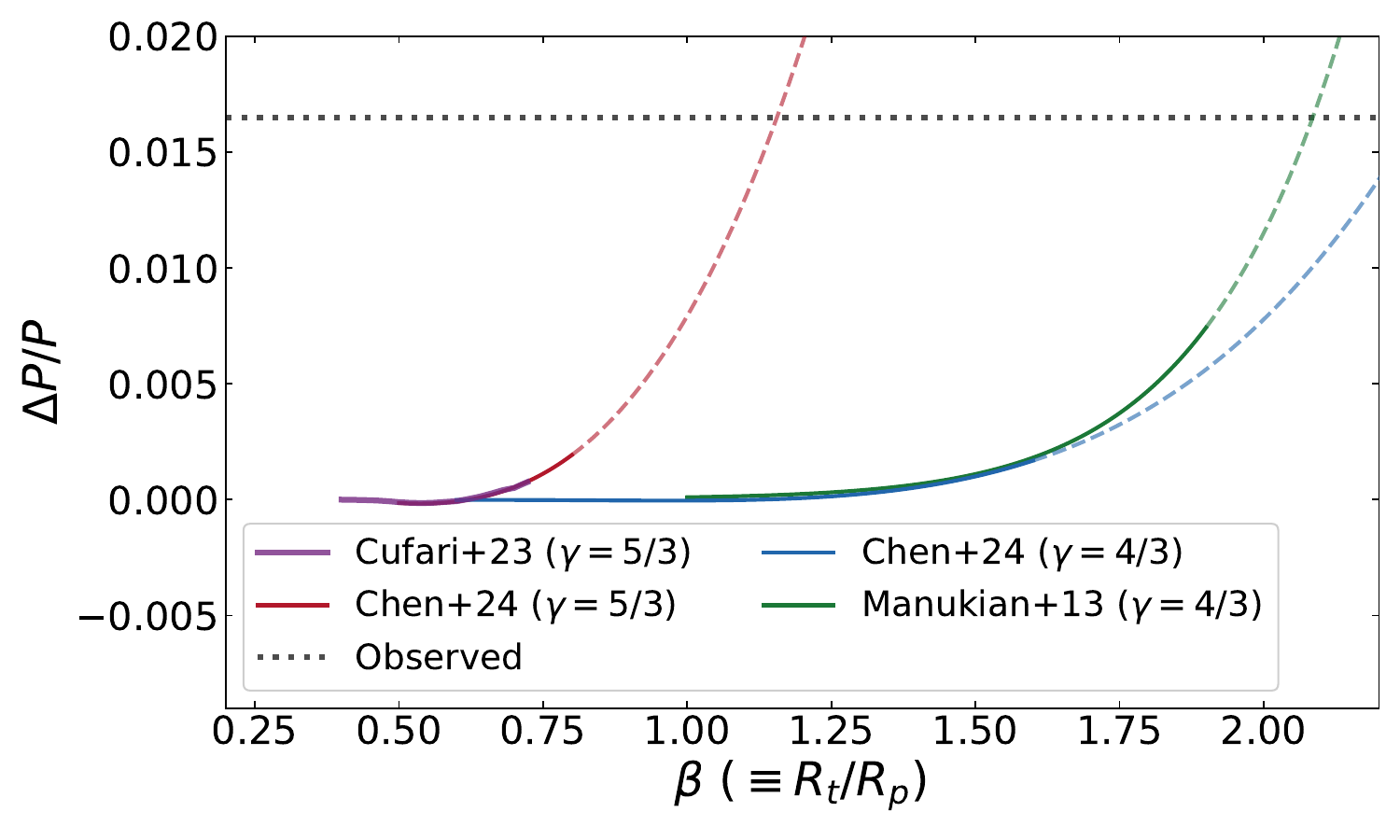}
    \caption{Magnitude of $\Delta P_{\rm orb}/P_{\rm orb}$ resulting from velocity kicks to the partial TDE remnant, as a function of the penetration factor from various prior works. The dashed lines indicate extrapolations beyond the simulation results of the respective studies.}
    \label{fig:pdot_vkick}
\end{figure}

\cite{Manukian13}, \cite{Gafton15}, and \cite{Chen24} found partial disruptions can impart velocity kicks up to $v_{\rm kick}\lesssim v_{\rm esc}$, the escape speed of the orbiting star ($=\sqrt{2GM_*/R_*}$), while \cite{Coughlin25} found $v_{\rm kick}$ can even exceed $v_{\rm esc}$. We relate an instantaneous kick velocity to a change in the specific orbital energy via $v_{\rm kick} = \sqrt{2\Delta \epsilon_{\rm kick}}$, following which the orbital period increases according to $\Delta P_{\rm orb}/P_{\rm orb} = (3/2) (\Delta a/a) = (3/2)(\Delta\epsilon/\epsilon)$. This can be cast in terms of $v_{\rm kick}/v_{\rm esc}$ as:
\begin{align}
    \frac{\Delta P_{\rm orb}}{P_{\rm orb}} \approx 6\times10^{-3}\; &\bigg(\frac{v_{\rm kick}}{v_{\rm esc}}\bigg)^2 \bigg(\frac{M_*}{M_\odot}\bigg) \bigg(\frac{R_*}{R_\odot}\bigg)^{-1} \nonumber \\
    &\times \bigg(\frac{M_\bullet}{10^6M_\odot}\bigg)^{-2/3}\bigg(\frac{P_{\rm orb}}{11\;\rm d}\bigg)^{2/3}
    \label{eq:vkick}
\end{align}
implying main sequence stars require $v_{\rm kick}\gtrsim v_{\rm esc}$ to attain the large $\dot{P}\sim10^{-2}$ observed in Ansky. 

This is difficult to reconcile with the finding that larger kicks occur for more deeply-plunging orbits (larger $\beta\equiv R_t/R_p$; e.g.~Figure 7 of \citealt{Gafton15}). It is generally found that for $\beta\lesssim 0.62$ the orbit shrinks, and only for $\beta\gtrsim 0.62$ can the velocity kick outweigh tidal dissipation to expand the orbit (see e.g.~Figure 1 of \citealt{Cufari23} and Figure 4 of \citealt{Chen24}). In Fig.~\ref{fig:pdot_vkick}, we show the fractional change in orbital period resulting from pericenter kicks as a function of $\beta$ resulting from several of these studies. We used the empirical fits of $\Delta\epsilon_{\rm orb}(\beta)$ provided in Eq. 9 of \cite{Chen24} and $v_{\rm kick}/v_{\rm esc}(\beta)$ in Fig. 3 of \cite{Manukian13}. The large magnitude of $\dot{P}$ is only attained for near-complete disruptions ($\beta\approx1$ for $\gamma=5/3$ and $\beta\approx2$ for $\gamma=4/3$), at which point $>20$ consecutive eruptions of near-constant luminosity become impossible to explain. It is thus unlikely that kicks resulting from tidal interactions can explain the period change.

One remaining possibility may be a change not in the orbital period itself, but the return time of the debris after successive pericenter passages. See for instance \cite{Bandopadhyay24}, who found the peak fallback time scales with the angular velocity of the stellar core ($\Omega$) as $\Big(1+2\Omega/\sqrt{GM_\bullet/r_p^3}\Big)^{0.8}$. A thorough investigation of this idea is beyond the scope of this work.

\subsection{General relativistic precession}  \label{subsec:gr_precession}

Given the relatively constant burst energetics through 2025, it is worth considering whether the observed $\dot{P}$ may be caused by an oscillatory change to the system geometry and/or light travel time due to relativistic precession(s), rather than a true change to the orbital semimajor axis. These are expected to induce measurable effects on QPE timings (e.g.~\citealt{Xian21,Linial23b, Linial24c,Franchini23,Zhou24a}), and such variations have already been observed in some cases \citep{Chakraborty24,Miniutti25,Arcodia26}. Models 2 and 4 of Section~\ref{sec:results} found a long-term precession period of $400-1050$ epochs ($11-27$ yr) and an amplitude of $1000-4500$ days; we consider whether the values can be reproduced.

\begin{figure}
    \centering
    \includegraphics[width=\linewidth]{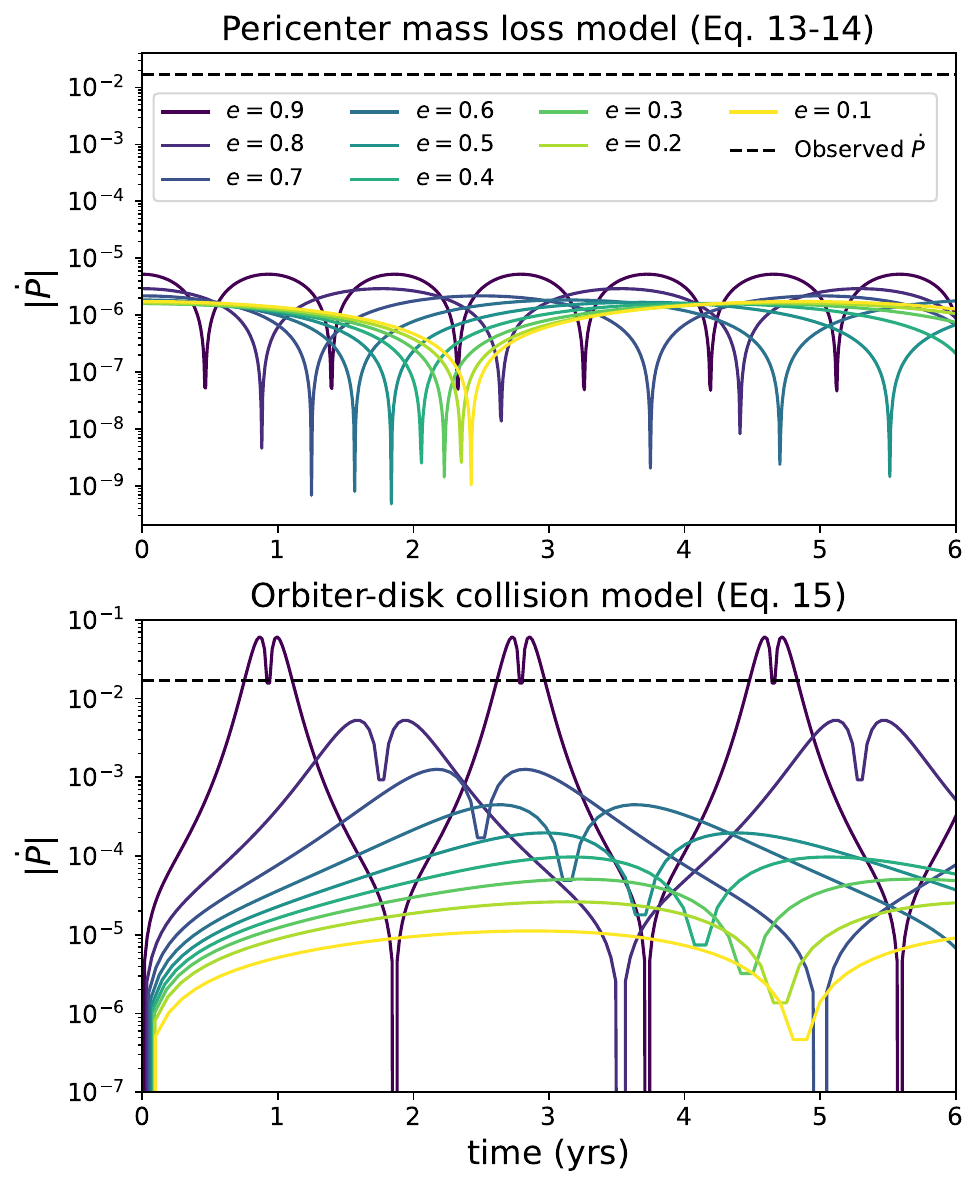}
    \caption{$|\dot{P}|$ for two general relativistic precession models, with fiducial values $M_\bullet=10^6M_\odot$ and $P_{\rm orb}=11$ days. \textbf{Top:} model assuming QPE timings are set by pericenter passages (Eq.~\ref{eq:gr_aps}-\ref{eq:gr_aps_amp}). \textbf{Bottom:} model assuming timings are set by orbiter-disk collisions (Eq.~\ref{eq:itai}; \citealt{Linial24a}).}
    \label{fig:precession}
\end{figure}

The simplest relativistic effect is Schwarzschild (apsidal) precession. In this case, one would observe a sinusoidal period modulation due to light travel delays over the apsidal precession timescale ($T_{\rm aps}$), which is approximately $T_{\rm aps}/P_{\rm orb} \approx a(1-e^2)/(3R_g)$. The semimajor axis is not known precisely due to uncertainty in $M_\bullet$, but $a$ has a characteristic value:
\begin{equation}
    a = 980 R_g\bigg(\frac{P_{\rm orb}}{11\;\rm d}\bigg)^{2/3}\bigg(\frac{M_\bullet}{10^6M_\odot}\bigg)^{-2/3}
    \label{eq:sma}
\end{equation}
meaning apsidal precession occurs over a timescale:
\begin{equation}
    T_{\rm aps} \approx 9.8\:\mathrm{yr}\:(1-e^2)\bigg(\frac{P_{\rm orb}}{11\;\rm{d}}\bigg)^{5/3}\bigg(\frac{M_\bullet}{10^6M_\odot}\bigg)^{-2/3}
    \label{eq:T_aps}
\end{equation}
Let $P_r$ be the radial period between successive pericenter passages, such that the $n$th passage is observed at:
\begin{equation}
    t_n \approx nP_{\rm orb} + \frac{r_p}{c}\cos\bigg(\frac{2\pi nP_{\rm orb}}{T_{\rm aps}}\bigg)
\end{equation}
where we have assumed that the emission is generated close to $r_p$. The period measured by a distant observer is $P=dt_n/dn$, so that the apparent $dP/dt$ is:
\begin{equation}
    \dot{P} = \frac{d^2t_n}{dn^2}\frac{dn}{dt} = - \frac{4\pi^2 r_pP_{\rm orb}}{cT_{\rm aps}^2}\cos \bigg(\frac{2\pi nP_{\rm orb}}{T_{\rm aps}}\bigg)
    \label{eq:gr_aps}
\end{equation}
where $dn/dt=1/P_{\rm orb}$. The maximum apparent period change resulting from this expression is:
\begin{equation}
    \dot{P}_{\rm max} \approx 1.9\times10^{-6}\;\frac{1-e}{(1-e^2)^2}\bigg(\frac{P_{\rm orb}}{11\;\rm d}\bigg)^{-5/3}\bigg(\frac{M_\bullet}{10^6M_\odot}\bigg)^{5/3}
    \label{eq:gr_aps_amp}
\end{equation}
which is four orders of magnitude too small (unless $e\rightarrow1$). We plot $|\dot{P}|$ for some choices of $e$ in the top panel of Fig.~\ref{fig:precession}.

We also consider the $\dot{P}$ in a model where the QPEs arise from star-disk collisions rather than pericenter mass-loss. In this case, the impact timings change as the orbiter and disk precess, so $\dot{P}$ reflects a true change in the system geometry, not just a light travel-time effect. The period derivative in this scenario was considered in Section 5.2, Eq.~40 of \cite{Linial24a}:
\begin{equation}
    \dot{P} \approx 10^{-4} \frac{e\sin\theta}{(1-e^2)(1+e\cos\theta)^3}\bigg(\frac{P_{\rm orb}}{11\;\rm d}\bigg)^{-4/3}\bigg(\frac{M_\bullet}{10^6M_\odot}\bigg)^{4/3}
    \label{eq:itai}
\end{equation}
For all but the most eccentric orbits, it remains difficult to produce the observed $\dot{P}\sim10^{-2}$ (Fig.~\ref{fig:precession} bottom). Even then, high-$e$ orbits maintain a large $\dot{P}$ only for a brief period of $\lesssim$few months due to their short $T_{\rm aps}$.

\begin{figure}
    \centering
    \includegraphics[width=\linewidth]{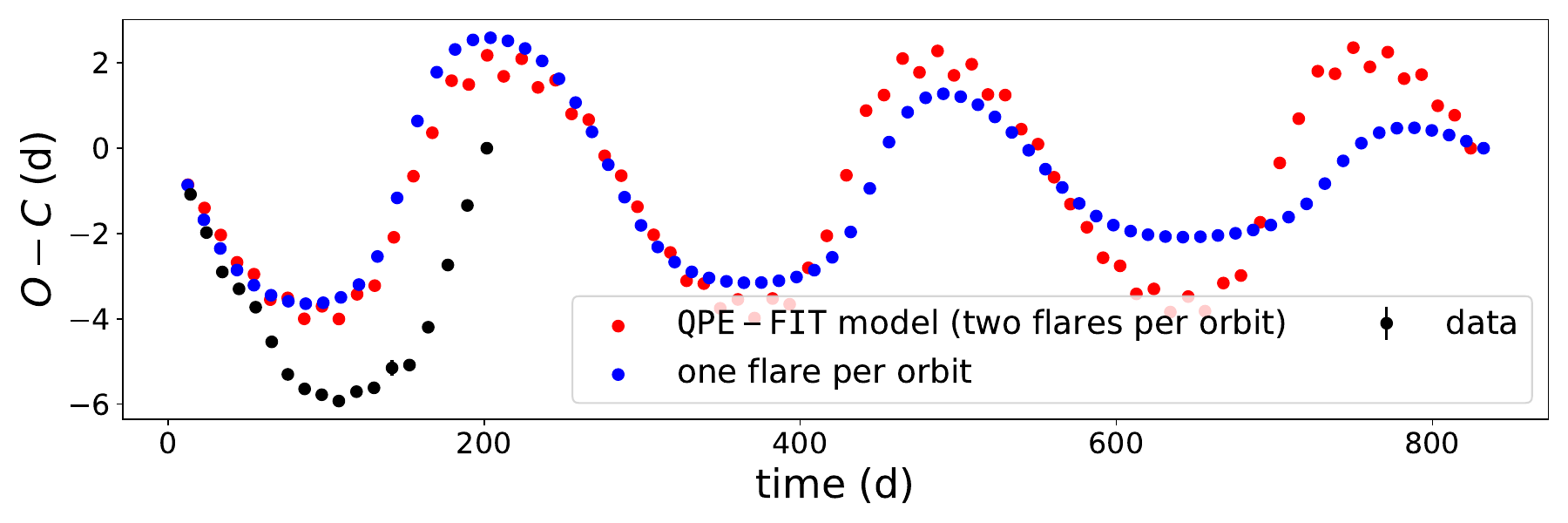}
    \caption{$O-C$ diagrams of data (black) and maximum-likelihood \texttt{QPE-FIT} models assuming two (red) or one (blue) flares are produced per orbit.}
    \label{fig:qpefit}
\end{figure}

Further super-orbital variations, with longer timescales and larger amplitude, can occur due to Lense-Thirring (nodal) precession of the orbiter and/or the accretion disk (e.g.~\citealt{Franchini23}). While this is more difficult to model analytically, the effect can be readily simulated with the open-source EMRI/disk trajectory modeling code \texttt{QPE-FIT\footnote{\href{https://github.com/joheenc/QPE-FIT}{https://github.com/joheenc/QPE-FIT}}} v0.1.11 \citep{Chakraborty25c}, which performs Bayesian inference using QPE timings to infer EMRI/disk parameters while including relativistic precession of the orbiter and disk. We show the best-fitting models in Fig.~\ref{fig:qpefit}: though the modulation amplitude can be approximately reproduced, the long modulation timescale cannot be captured simultaneously. This is the same tension we saw in Eq.~\ref{eq:itai}: large-amplitude geometric delays require highly eccentric orbits, but larger $e$ shortens the precession timescales compared to the year baseline over which we observe a stable $\dot{P}$. 

It is worth noting the \textit{short}-period oscillation in $O-C$ models 3-4, with $P_s\approx155$ d and amplitude $0.8$ d, is plausibly consistent with relativistic precession. A 155 d period requires $e\gtrsim0.97$ (Eq.~\ref{eq:T_aps})---which is not impossible (e.g.~the orbits of S-stars around Sgr A*, \citealt{Burkert24})---and the amplitude $\ll T_{\rm rec}$ is significantly easier to obtain.

In any case, from the above arguments and the $O-C$ constraints it appears unlikely that the apparent $\dot{P}$ is entirely the result of relativistic precession, due to the long modulation timescales ($11-27$ yr) and amplitudes ($1000-4500$ days). It is possible that a large SMBH mass $\gg 10^6M_\bullet$ may increase the variation amplitude (e.g.~Eq.~\ref{eq:itai}), but this would both decrease the precession period (Eq.~\ref{eq:T_aps}) and be in tension with host galaxy scaling relation-based estimates of $M_\bullet$ \citep{Sanchez24,Zhu25}. It may thus be that reproducing the data requires a previously unconsidered source of precession-induced timing variation---an increasingly common conclusion of QPE timing studies \citep{Miniutti25,Arcodia26}. While the most direct confirmation of this model would be eventually measuring the change to $\dot{P}$ predicted by long-term precession after a few years (Fig.~\ref{fig:evol}), this may be achieved earlier by directly measuring $\ddot{P}$.

\subsection{Hierarchical SMBH binary} \label{subsec:third_body}

Another possibility is the presence of an outer SMBH binary which induces the apparent period change via geometric light travel-time delays due to reflex motion of the QPE-emitting system, a situation comparable to models 2 and 4 of Section~\ref{sec:results} (see also \citealt{Miniutti25}). An SMBH binary with period $P_{\rm SMBH}$, orbital velocity $v_{\rm orb}$, and frequency $\omega_{\rm orb}$ would induce an apparent $\dot{P}=(P\omega_{\rm orb}v_{\rm orb}/c)\cos(\omega_{\rm orb}t)$ for an edge-on observer. For an equal-mass circular binary, the maximum apparent derivative for an edge-on observer is then:
\begin{equation}
    \dot{P} \approx 4.4\times10^{-5}\;\bigg(\frac{M_{\rm tot}}{2\times10^6 M_\odot}\bigg)^{1/3} \bigg(\frac{P_{\rm SMBH}}{20\;\mathrm{yr}}\bigg)^{-4/3} \bigg(\frac{P}{11\;\mathrm{d}}\bigg)
    \label{eq:smbhb}
\end{equation} \\
with $\dot{P}$ decreased by $\sin\theta$ for smaller viewing angles. An SMBH binary thus cannot account for the period change due to our simultaneous constraints on the large magnitude of $\dot{P}$ and the long oscillation timescale ($\gtrsim 11$ yr).

\subsection{Disk instability model} \label{subsec:disk_instability}

Prior work has also explored the applicability of disk instability models for QPEs (e.g.~\citealt{Sniegowska20,Pan22}). As in the standard \cite{SS73} treatment, they find that SMBH accretion disks of sufficiently high Eddington ratio ($\dot{m}\equiv\dot{M}/\dot{M}_{\rm Edd}\gtrsim 0.01$) are thermally and viscously unstable in the inner radiation pressure-dominated region. These disks are known to undergo limit cycles, whereby material gradually accumulates, then rapidly accretes onto the black hole in recurrent outbursts. While this normally occurs on the viscous timescale, for narrow unstable regions---of width $\Delta R$ at radius $R$---the timescale can be reduced by order $\Delta R/R$ \citep{Sniegowska20}. \cite{Pan22} proposed that large-scale poloidal magnetic fields can confine the unstable region to $\Delta R\sim 0.1R_S$ by driving winds to remove angular momentum and cool the disk; QPEs are then driven by limit cycles in this inner ring rather than the entire disk, producing qualitative agreement with the timescales and energetics of most QPEs \citep{Pan23}.

\cite{Pan25} subsequently found that burst recurrence times may vary by factors of several, even for percent-level changes in $\dot{m}$. They identified a critical threshold near $\dot{m}_{\rm crit}\approx 0.1$, dependent on the magnetic field strength, below which $T_{\rm rec}$ remains relatively stable and above which the sensitivity of $T_{\rm rec}$ to $\dot{m}$ increases dramatically (see their Figure 1). It is thus possible that, if $\dot{m}$ increased from $\lesssim \dot{m}_{\rm crit}$ in 2024 to $\gtrsim \dot{m}_{\rm crit}$ in 2025, and is now increasing steadily, the observed period evolution in Ansky could be produced. Only a modest long-term increase in $\dot{m}$ would be required, and such a trend has already been observed in other QPEs (e.g.~\citealt{Miniutti23a}).

There are some immediate uncertainties in this picture. Ansky's quiescence emission was detected in three \textit{XMM-Newton} observations: OBSID 0935191501 in July 2024 (presented in \citealt{Hernandez25a}), and OBSIDs 096450101 and 096450301 in July 2025. Assuming a multicolor disk blackbody model for both epochs, we estimated 0.3-2 keV luminosities of $L_{q,2024}=(4.1\pm 1.2)\times 10^{40}$ and $L_{q,2025}=(2.9\pm 0.3)\times 10^{40}$ erg s$^{-1}$, respectively, i.e. a constant luminosity within errors. The steady-state accretion rate is then:
\begin{equation}
    \dot{m} \approx 0.03 \bigg(\frac{L_{X,\rm{quiesc}}}{3\times10^{40}\rm{erg}\;\rm{s}^{-1}}\bigg)\bigg(\frac{\eta_X}{100}\bigg) \bigg(\frac{M_\bullet}{10^6M_\odot}\bigg)
\end{equation}
where $\eta_X\equiv L_{\rm bol}/L_X$ is the bolometric correction factor. Achieving $\dot{m}_{\rm crit}\approx0.1$ would require $\eta_X\gtrsim 300$, a factor of a $3-30$ larger than typical for TDE disks (e.g.~\citealt{Mummery23}). While this is not impossible given uncertainties in $M_\bullet$, $\eta_X$, and $\dot{m}_{\rm crit}$, further work is needed to determine whether these tensions present an existential problem for the disk instability model.

\section{Conclusion} \label{sec:conclusion}

We reported the surprising result that the quasi-period eruptions (QPEs) in ZTF19acnskyy/``Ansky'' have shown a smoothly \textit{increasing} period---exactly opposite to the gravitational wave and/or viscous drag-driven inspiral predicted by EMRI models---with a time-derivative $\dot{P}\approx(1.7\pm0.02)\times10^{-2}$ throughout 2025-2026 (Fig.~\ref{fig:oc}). Backwards-extrapolating the $\dot{P}$ implies that the $4.5-7$ day recurrence times observed in 2024 data are expected $\gtrsim 20$ QPEs before those in 2025, providing a partial explanation for the previous data. The period derivative has remained constant for over a full year thus far, ruling out period oscillations with $P\lesssim 11$ yr or amplitude $\lesssim 1000$ d. It remains possible that a very long-term sinusoidal modulation, of period $\sim 11-27$ yr and amplitude $\sim1000-4500$ d, is responsible for the observed period increase, though that is about equally challenging to explain. 

We considered several physical models for the period derivative, all of which face shortcomings:
\begin{enumerate}
    \item Prior work has calculated the orbital evolution due to pericenter mass-transfer in eccentric stellar-mass binaries via the orbit-averaged Lagrange planetary equations (e.g.~\citealt{Sepinsky09,Dosopoulou16b}). Extending these results to EMRIs, the observed $\dot{P}$ can be reproduced for fractional mass losses of $\Delta M_*\sim10^{-3}M_*$ per passage (Eq.~\ref{eq:delta_P}), even for non-conservative mass-transfer ($\dot{M}_{\rm tot},\dot{J}_{\rm orb}\neq0$). Three uncertainties are whether the stellar structure can sustain many consecutive comparable-energy bursts after losing $\sim$10\% of its mass over a year (Fig.~\ref{fig:evol}); whether the $\sim$58
    day viscous time at pericenter can be reconciled with the $\lesssim3$-day burst duration; and the extent to which L2 mass loss modifies the expected period evolution due to mass-transfer.
    \item Hydrodynamical simulations of partial TDEs have found that for large penetration factors, the remnants can receive a positive-energy kicks from asymmetric mass-loss and/or the reformation of the core, making their orbit progressively unbound \citep{Manukian13,Gafton15,Cufari23,Chen24,Coughlin25}. The observed magnitude of $\dot{P}$ would require $v_{\rm kick}\gtrsim v_{\rm esc}$ (Eq.~\ref{eq:vkick}), which can be obtained for relatively deeply-plunging orbits ($R_p\approx R_t$). However, this is impossible to reconcile with the energetics across $>30$ eruptions.
    \item We also examined whether the period changes may be produced by general relativistic precession. We considered the effects of apsidal motion (Eqns.~\ref{eq:gr_aps}-\ref{eq:itai}; \citealt{Linial24a}) and nodal precession of the orbiter/accretion disk \citep{Franchini23,Chakraborty25c}, finding the large $\dot{P}$ amplitude and the long duration over which it is stable are difficult to reproduce simultaneously (Figs.~\ref{fig:precession}-\ref{fig:qpefit}). Given that no known combination of relativistic precessions appears readily able to explain the data, models for precession-induced timing variations may require extension, a theme also seen in other recent QPE timing studies \citep{Miniutti25,Arcodia26}.
    \item In a similar vein, we considered whether an outer SMBH binary can produce an apparent period change due to light travel-time delays induced by the reflex motion of the QPE-emitting system, and found the inferred $\dot{P}$ falls short by $\sim 3$ orders of magnitude.
    \item Disk instability models invoking radiation pressure-driven limit cycles in a magnetically-confined ring near the ISCO (e.g.~\citealt{Sniegowska20,Pan22}) appear qualitatively able to reproduce both the relatively steady $4.5-7$ day recurrence times in 2024 and the smooth increase in 2025, if the accretion rate steadily increases by a few percent \citep{Pan25}. However, the measured quiescence X-ray luminosity and recurrence timescales are not immediately compatible with the numerical ranges explored in the literature, motivating further work. Moreover, it remains unclear what physical process can set such a stable underlying clock in the disk instability picture.
\end{enumerate}

Most of the models predict an eventual change to the QPE properties. Scenarios involving a secular increase to the EMRI orbital radius require large stellar mass loss to drive the dramatic period change, implying both that the burst luminosities should change as the stellar structure responds, and that the QPEs should have a finite duration of $\lesssim$years. The relativistic precession model clearly predicts that eventually $\dot{P}$ should decrease in magnitude, then change sign, though the constraints already placed from the long-term stability of $\dot{P}$ mean this would only be observable after at least a few years (Fig.~\ref{fig:evol}). Continued monitoring of the burst timings and energetics, and continued refinement of physical models for QPEs, are thus well-motivated as the remarkable diversity of unexpected QPE phenomenology only grows.

\section*{Acknowledgements}
We thank Keith Gendreau, Zaven Arzoumanian, Scott Hughes, Eric Coughlin, Xin Pan, Daniel D'Orazio, Hengxiao Guo, and Zhen Yan for discussions and data which contributed to this work. We thank the anonymous referee for constructive comments which improved the manuscript. We are particularly grateful to the \textit{NICER} team for providing an exceptionally high-quality dataset which uniquely enabled this analysis. GM acknowledges support from grant n. PID2023-147338NB-C21 funded by Spanish MICIU/AEI/10.13039/501100011033 and ERDF/EU. JCu acknowledges financial support from ANID -- FONDECYT Regular 1251444, and Millennium Science Initiative Program NCN$2023\_002$. LHG acknowledges financial support from ANID program FONDECYT Iniciaci\'on 11241477. MG is funded by Spanish MICIU/AEI/10.13039/501100011033 and ERDF/EU grant PID2023-147338NB-C21. The version of \texttt{QPE-FIT} (v0.1.11) used in this work can be found at the static repository \href{https://zenodo.org/records/18943053}{doi:10.5281/zenodo.18943053}.

\bibliography{refs}{}
\bibliographystyle{aasjournal}

\appendix
\section{On the interpretation of residuals in $O-C$ curves} \label{app:oc_noise}

The interpretation of the $O-C$ residuals depends on whether the processes generating QPE timings are governed by a deterministic underlying clock (e.g.~a binary orbit) or not (e.g.~an accretion instability). Suppose there is an underlying ``noisy clock'', with mean period $P_0$ and derivative $\dot{P}$, that exhibits uncorrelated random timing noise of characteristic size $\Delta t$. Then, the $n$th burst occurs at:
\begin{equation}
    t_n = P_0n + \frac{1}{2}P_0\dot{P}n^2 + \Delta t_n
\end{equation}
The $O-C$ residuals after subtracting the deterministic model $P_0n+(1/2)P_0\dot{P}n^2$ are: 
\begin{equation}
(O-C)_{n,\rm clock} = \Delta t_n
\end{equation}
i.e. uncorrelated white noise. However, if no underlying clock exists, each burst timing depends on the previous one:
\begin{align}
    t_n &= t_{n-1} + P_0 + P_0\dot{P}n + \Delta t_n \nonumber\\
    &\approx P_0n+\frac{1}{2}P_0\dot{P}n^2 + \sum_{m=1}^n \Delta t_m
\end{align}
yielding cumulative residuals:
\begin{equation}
    (O-C)_{n,\rm no\;clock} = \sum_{m=1}^{n}\Delta t_m    
\end{equation}
which is a random walk with a red noise power spectrum $\propto 1/f^2$. The characteristic amplitude of a spurious sinusoidal signal over the full baseline is thus of order:
\begin{equation}
    A_{\rm rms} \approx \Delta t\sqrt{N}
    \label{eq:A_rms}
\end{equation}
For a sinusoid of period $P_{\rm sin}<N$, the RMS displacement shrinks to $\Delta t\sqrt{P_{\rm sin}}$, meaning the longest periods allowed by the observation---that is, the full baseline---are the most prone to the appearance of false sine-like residuals arising from random walk noise. In other words, when interpreting residuals of QPE timing $O-C$ curves \textit{without} the EMRI picture, one must account for the possibility that apparent periodicities of modest amplitudes and periods comparable to the observing baseline are driven by stochastic noise---an insight all too familiar in the contexts of of accretion variability \citep{Vaughan16} and even binary eclipse timing \citep{Koen06}.

Consider the two $O-C$ components identified in Section~\ref{sec:results}. Prior to including the period derivative term, the $O-C$ curve reaches a maximum of $\sim 30$ days over 33 epochs (Fig.~\ref{fig:oc}). Explaining this via random noise requires $\Delta t\sim 5.3$ d (Eq.~\ref{eq:A_rms}). Compared to the underlying periodicity of $\sim 11$ d, this demands a clock unstable at nearly the 50\% level. This is certainly unreasonable for any orbital phenomenon; while such a statement cannot be made definitively in the disk instability model, it is nevertheless an extreme variation. The physical reality of the $\dot{P}$ term is thus almost model-independent.

On the other hand, for the short-period oscillation with $A_s=0.8$ d, Eq.~\ref{eq:A_rms} gives $\Delta t\sim 0.14$ d, only $\sim1$\% of the QPE recurrence time. The deterministic orbital model predicts white noise in the $O-C$ residuals, so a sine term to capture the additional structure is well-justified even for the small amplitude (i.e. models 3-4 of Section~\ref{sec:results}). However, disk instability models could easily produce red noise at this level; if the true QPE-generating process is not orbital, then models 3-4 are not necessary. While it is a subtle point, the necessity of additional terms to capture structured residuals in the $O-C$ curve depends on the physical interpretation of what one believes is actually driving the QPEs.

\end{document}